\documentclass[preprint]{revtex4-1}
\usepackage{graphicx}
\usepackage{subfig}

\usepackage{graphicx}
\usepackage{dcolumn}
\usepackage{bm}

\begin{document}

\title{Random Sequential Adsorption on Fractals}

\author{Michal Ciesla$^{1}$}
 \email{michal.ciesla@uj.edu.pl}
\author{Jakub Barbasz$^{1,2}$}
 \email{ncbarbas@cyf-kr.edu.pl}

\affiliation{
$^1$ M. Smoluchowski Institute of Physics, Jagiellonian University, 30-059 Kraków, Reymonta 4, Poland. \\
$^2$ Institute of Catalysis and Surface Chemistry, Polish Academy of Sciences, 30-239 Kraków, Niezapominajek 8, Poland.
}
\date{\today}

\begin{abstract}
Irreversible adsorption of spheres on flat collectors having dimension $d<2$ is studied. Molecules are adsorbed on Sierpinski’s Triangle and Carpet like fractals ($1<d<2$), and on General Cantor Set ($d<1$). Adsorption process is modeled numerically using Random Sequential Adsorption (RSA) algorithm. The paper concentrates on measurement of fundamental properties of coverages, i.e. maximal random coverage ratio and density autocorrelation function, as well as RSA kinetics. Obtained results allow to improve phenomenological relation between maximal random coverage ratio and collector dimension. Moreover, simulations show that, in general, most of known dimensional properties of adsorbed monolayers are valid for non-integer dimensions.

\end{abstract}

\maketitle

\section{Introduction}
Most of the effort in the research on irreversible adsorption is focused on flat homogeneous surfaces as it can be further directly exploited in material sciences. However, experiments using fractal-like collectors were also made \cite{bib:Kinge2008}, though preparation of such surfaces is complicated. On the other hand, there are many porous media in nature and adsorption may play an important role there \cite{bib:Pfeifer1983, bib:Avnir1983}. For example,  coral fractal-like structure helps them to catch plankton effectively \cite{bib:Basillais1998}. Adsorption on fractal collectors might also be applied in environmental protection in designing effective water or air filters \cite{bib:Khasanov1991}.
\par
Hard spheres packing has a rich research history, both in physics and maths. The greatest interest has been shown in the maximal possible packing \cite{bib:Conway1998, bib:Rogers1964}. The most common property of a given packing is its density. In $d$ dimensional Euclidean space, it is given by
\begin{equation}
\theta_{max}(d) = \rho \cdot v(d) 
\end{equation}
where $\rho$ is a number of spheres in a unit volume, and
\begin{equation}
v(d) = \frac{ \pi^{d/2} }{ \Gamma (1+d/2) } \,\, r_0^d 
\end{equation}
is a volume of $d$ dimensional sphere having radius $r_0$. $\Gamma(x)$ is the Euler gamma function.
\par
Presented work focuses on the properties of maximal random coverages, built of spherical particles on flat collectors having fractal dimension of a non-integer value. Earlier, similar studies concentrated mostly on adsorption on lattice collectors. Their popularity was powered by a number of analytic results obtained, e.g.~\cite{bib:Privman2004, bib:Fan1991}. Other investigations concerning fractal collectors concentrate on diffusion properties \cite{bib:Nazzarro1996, bib:Loscar2003} or adsorption on rough surfaces \cite{bib:Cole1986, bib:Pfeifer1989}. Fractal structures formed in the adsorption processes were also observed \cite{bib:Brilliantov1998}.
\par
Dimensional properties of adsorption layers were studied mostly for collectors having integer dimensions, with special attention paid to two-dimensional case, e.g. \cite{bib:Feder1980, bib:Swendsen1981, bib:Privman1991, bib:Torquato2006}, due to its potential application in chemistry and material science. Here, the only analytically solved case is a one-dimensional problem known also as ”car parking problem” for which $\theta_{max}(1)=0.748...$ \cite{bib:Renyi1963}. The analysis of hard spheres random packing for $2 \le d \le 6$, nicely reviewed in \cite{bib:Torquato2006}, shows that maximal random coverage ratio decreases with the growth of collector’s dimension:
\begin{equation}
\theta_{max}(d) = \frac{c_1 + c_2 d}{2^d}
\label{eq:fit1}
\end{equation}
with $c_1=0.202$ and $c_2=0.974$ \cite{bib:Torquato2002, bib:Torquato2006}. 
There are also some characteristics, which do not depend qualitatively on dimension.
For example the two-point density correlation function for maximal random coverages is known to have superexponential decay for large distances \cite{bib:Bonnier1994}
\begin{equation}
C(r) \sim \frac{1}{ \Gamma(r) } \left( \frac{2}{\ln \left( r/(2r_0) -1 \right)} \right)^{(r/2r_0)-1} \mbox{ for } r \to \infty, 
\label{eq:corfast}
\end{equation}
and logarithmic singularity when particles are in touch \cite{bib:Swendsen1981, bib:Privman1991}
\begin{equation}
C(r) \sim \ln \left( \frac{r}{2r_0} -1 \right) \mbox{ for } r \to (2r_0)^+ \mbox{ and } \theta \to \theta_{max}.
\label{eq:corlog}
\end{equation}
\par
The aim of this work is to check if the mentioned characteristics are valid for adsorption on collectors having non-integer fractal dimensions. The maximal coverages has been obtained here numerically by using Random Sequential Adsorption (RSA) algorithm \cite{bib:Feder1980} described in detail in Section \ref{sec:Model}. The kinetics of RSA simulation has also been of interest to the authors of this work, as it affects estimation of  $\theta_{max}(d)$. For sufficiently long simulation time the coverage ratio $\theta(t)$ scales with collector's dimension \cite{bib:Feder1980, bib:Swendsen1981}:
\begin{equation}
\theta_{max} - \theta(t) \sim t^{-1/d}.
\label{eq:feder}
\end{equation}
Here $\theta_{max} \equiv \theta(t \to \infty)$. Importance of the above relation is increased by the fact that it is also valid for other molecules, more complex than simple spheres \cite{bib:Ciesla2012, bib:Adamczyk2010}. To model collectors having dimension from a range of $d \in (1, 2)$ the Sierpinski's Triangle and, Carpet-like fractals are used. Adsorption for $d<1$ is studied using General Cantor Set. 
\section{Model}
\label{sec:Model}
\begin{figure}[htb]
\vspace{1cm}
\centerline{%
\includegraphics[width=8cm]{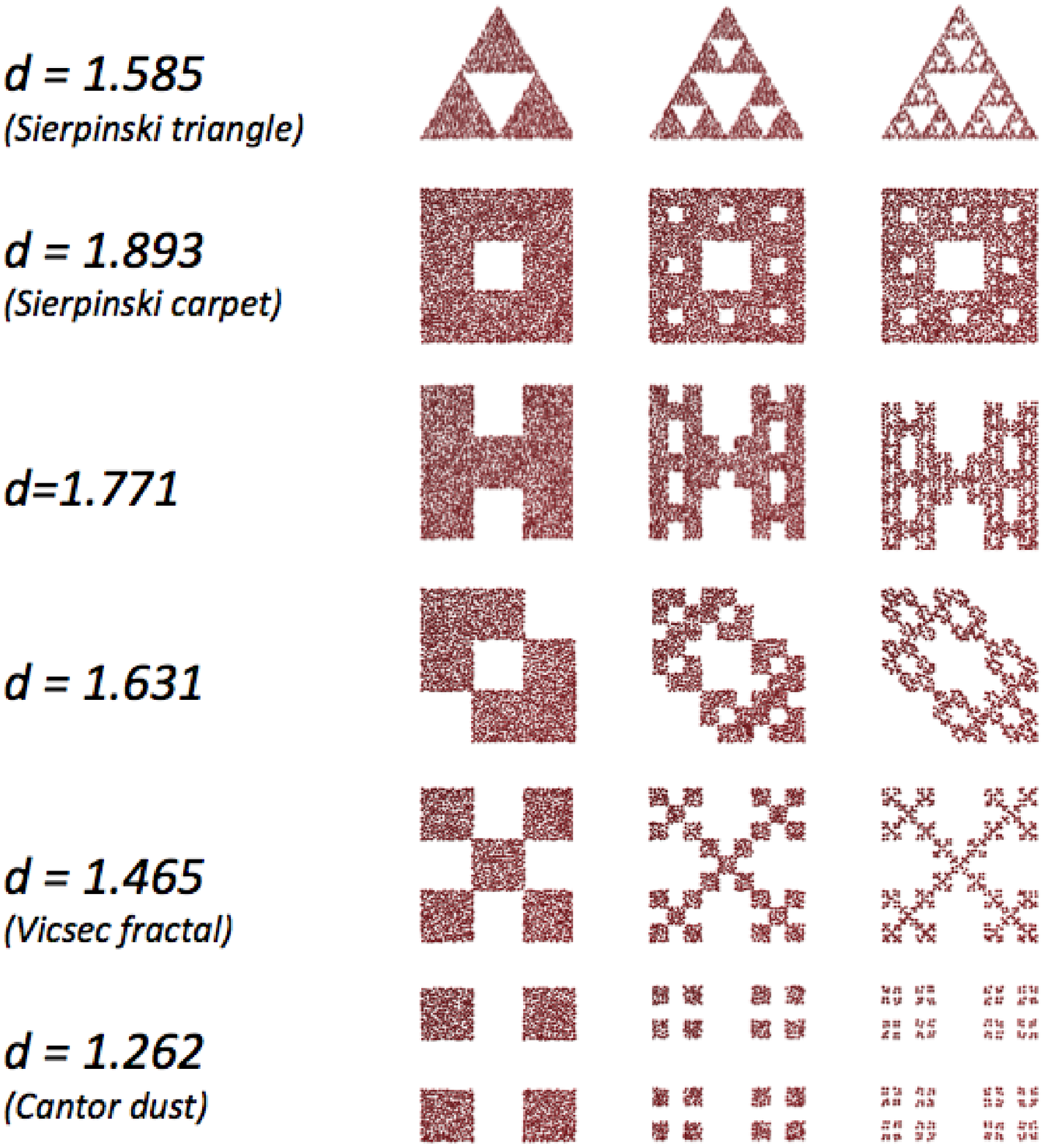}}
\caption{Maximal random coverages on first, second and sixth iteration of example fractals. The side length of square containing the fractal is 100$r$.}
\label{fig:examples}
\end{figure}

Maximal random coverages are generated using RSA algorithm, which has been successfully applied to study colloidal systems. It is based on independent, repeated attempts of adding sphere to a covering film. The numerical procedure is carried out in the following steps:
\begin{description}
\item[i] a virtual sphere is created with its centre position on a collector chosen randomly according to the uniform probability distribution;
\item[ii] an overlapping test is performed for previously adsorbed nearest neighbours of the virtual particle. The test checks if surface-to-surface distance between spheres is greater than zero;
\item[iii] if there is no overlap the virtual particle is irreversibly adsorbed and added to an existing covering layer. Its position does not change during further calculations;
\item[iiii] if there is an overlap the virtual sphere is removed and abandoned.
\end{description}
Attempts are repeated iteratively. Their number is usually expressed in a~dimensionless time units:
\begin{equation}
t_0 = N\frac{S_D}{S_C}
\end{equation}
where $N$ is a number of attempts and $S_D$ denotes an area of a single sphere projection on a collector. Typically, $S_C$ is a collector area.  Although formally the area of any fractal having $D<2$ is zero, $S_C$ denotes here an area of an overlapping square and is equal to $(100r)^2$, where $r$ is a radius of an adsorbed sphere. Collectors are modeled as a successive iteration of a given fractal. For each collector type at least $20$ independent numerical experiments have been performed. Sample coverages are presented in Fig.\ref{fig:examples}
Dimension of a given fractal is estimated using relation
\begin{equation}
d = \frac{\ln(p)}{\ln(s)},
\end{equation}
where $p$ is a number of identical smaller parts created in a single iteration and $s$ is a scale change. In case of Cantor Dust, for example, after one iteration four smaller squares appear after one iteration $(p=4)$ with a side length three times smaller than in the original square $(s=3)$. Therefore, dimension of Cantor Dust is $\ln(4) / \ln(3) \approx 1.262$
\par 
Coverage ratio of a given adsorption layer is determined by using random sampling of collector's points and checking whether they are covered by any, previously adsorbed sphere:
\begin{equation}  
\label{eq:qt}
\theta(t) = \frac{n_c(t)}{n}
\end{equation}
where $n_c(t)$ is a number of covered points after a dimensionless time $t$; $n$ denotes total number of sampled points. Here $n=10^6$, which provides statistical error at the level of $0.1\%$.
\section{Results and Discussion}
\subsection{Kinetics of the RSA}
There are two factors affecting maximal random coverage ratio obtained during simulation. First, the collector used is not an actual fractal but only a two-dimensional approximation and the number of adsorbed spheres clearly depends on the level of this approximation (see Fig.\ref{fig:examples}). Second, the maximal random coverage is defined for a collector having infinite size and is achieved after an infinite simulation time. 
\par
\begin{figure}[htb]
\vspace{1cm}
\centerline{%
\includegraphics[width=8cm]{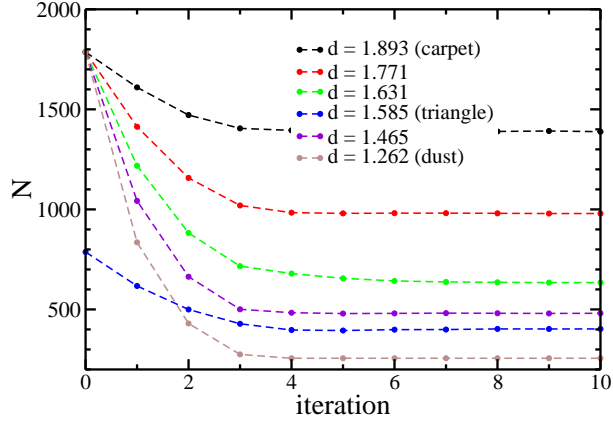}}
\caption{The number of adsorbed particles depends on a fractal's iteration. Iteration zero is a~solid triangle in case of the Sierpinski Triangle and a square in all the other cases.}
\label{fig:2dni}
\end{figure}
\begin{figure}[htb]
\vspace{1cm}
\centerline{%
\includegraphics[width=8cm]{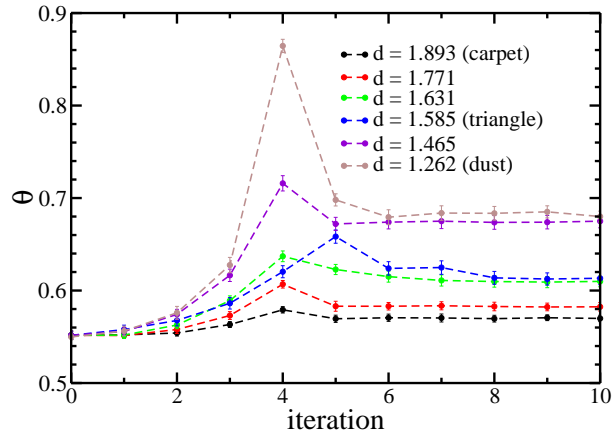}}
\caption{Coverage ratio after $t=10^5$ as a function of fractal's iteration. Iteration zero is a~solid triangle in case of the Sierpinski's Triangle and a square in all the other cases.}
\label{fig:2dti}
\end{figure}

To handle the first factor, the successive iterations of fractals have been used. The number of adsorbed particles becomes stable at about the $5^{th}$ fractal iteration (see Fig.\ref{fig:2dni}). Similarly, the coverage calculated using Eq.(\ref{eq:qt}) does not change thereafter, too (Fig.\ref{fig:2dti}). Therefore, using the $10^{th}$ iteration has been assumed to be save and to give reliable results.
\par
\begin{figure}[htb]
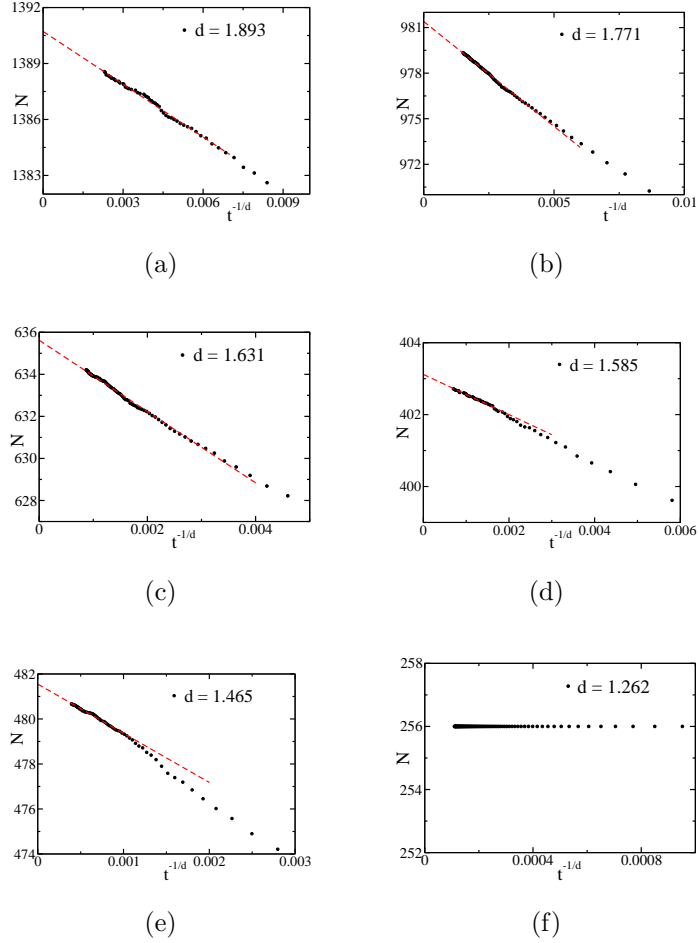

\vspace{0.1cm}
\centerline{%
\subfloat[]{\includegraphics[width=4cm]{fd495}}
\hspace{1cm}
\subfloat[]{\includegraphics[width=4cm]{fd381}}
}
\vspace{0.1cm}
\centerline{%
\subfloat[]{\includegraphics[width=4cm]{fd427}}
\hspace{1cm}
\subfloat[]{\includegraphics[width=4cm]{fdt}}
}
\vspace{0.3cm}
\centerline{%
\subfloat[]{\includegraphics[width=4cm]{fd341}}
\hspace{1cm}
\subfloat[]{\includegraphics[width=4cm]{fd325}}
}
\caption{Asymptotic character of RSA kinetics for collectors of different dimensions. The fits correspond to Eq.(\ref{eq:feder}). (a)~$d=1.893$ (Sierpinski Carpet), (b)~$d=1.771$, (c)~$d=1.631$, (d)~$d=1.585$ (Sierpinski Triangle), (e)~$d=1.465$ (Vicsec fractal), (f)~$d=1.262$ (Cantor Dust).}
\label{fig:federlaw}
\end{figure}
\begin{figure}[htb]
\vspace{1cm}
\centerline{%
\includegraphics[width=8cm]{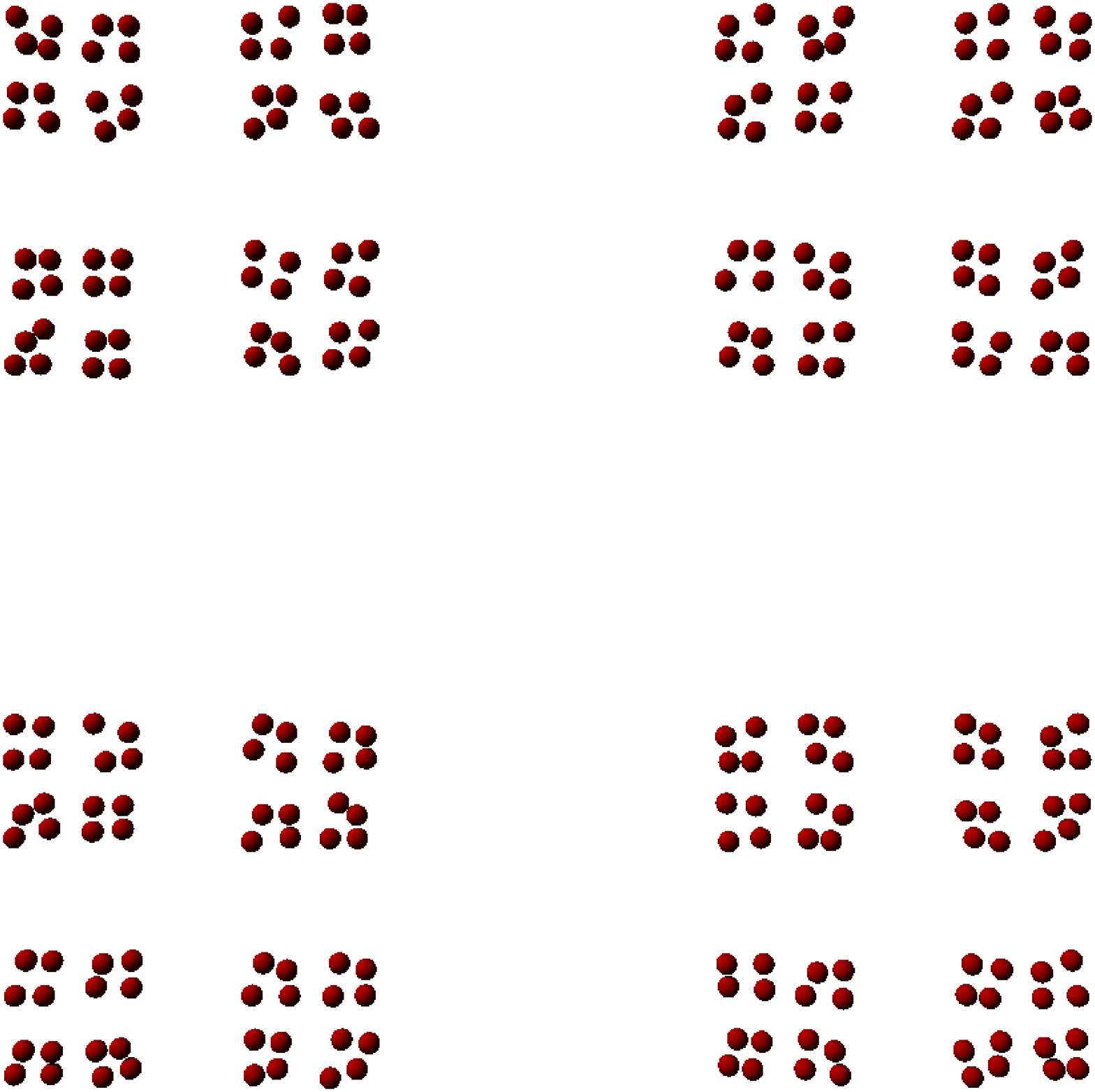}}
\caption{Sample coverage on the Cantor Dust collector.}
\label{fig:cantordust}
\end{figure}

To obtain maximal random coverage ratio after an infinite time, the RSA kinetics model has to be used. The analysis of numerical experiments shows that RSA for spheres in integer dimensions obeys the Feder's law (\ref{eq:feder}). The results presented in Fig.\ref{fig:federlaw} confirm that for almost all the cases analysed in this study. Additionally, the difference between observed maximal number of adsorbed particles for $t=10^5$ and the asymptotic value does not exceed $0.2\%$.
Coverage does not grow asymptotically only in case of adsorption on the sole incoherent collector studied, the Cantor Dust. In this case, the maximal possible number of $256$ particles  is adsorbed after relatively short time. This is due to geometrical relations between collector size and spheres' diameter (see Fig.\ref{fig:cantordust}).
After four iterations, each of the remaining squares has a side length of $100/3^4 r_0 \approx 1.234 r_0$ and can adsorb only one sphere. Moreover, due to the large distance between parts of the fractal, particle adsorbed on one square cannot block absorption events on its neighbouring squares. Therefore, each square contains a single sphere. It is noteworthy that further iterations of the fractal do not affect the above reasoning. Therefore, after catching $4^4=256$ spheres no further adsorption acts are possible.
\subsection{The Maximal Random Coverage Ratio}
\begin{figure}[htb]
\vspace{1cm}
\centerline{%
\includegraphics[width=8cm]{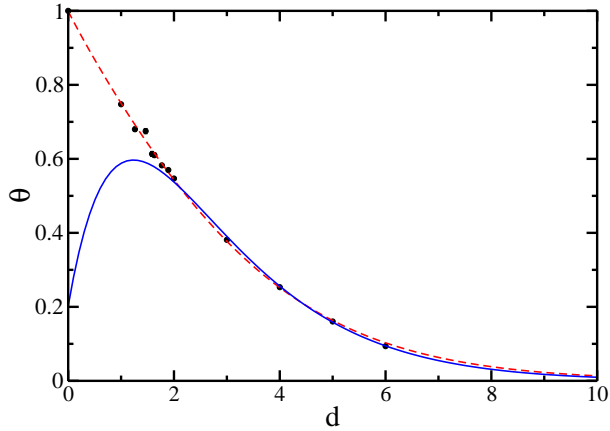}}
\caption{Maximal random coverage ratio in different dimensions. Dots represents obtained data, solid line is the (\ref{eq:fit1}) fit ($c_1=0.978$ and $c_2=0.581$), whereas dashed line is the (\ref{eq:fit2}) fit with $c_1=0.999$, $c_2=0.416$ and $c_3=0.086$.}
\label{fig:qd16}
\end{figure}

Investigating the relationship between maximal random coverage ratio and fractal dimension is the main purpose of this study. Although, there are number of studies providing lower and upper limits for a saturated coverage, e.g. \cite{bib:Ball1992}, the best direct fit so far is given by Eq.(\ref{eq:fit1}). It perfectly matches simulation data for $2 \le d \le 6$ \cite{bib:Torquato2006}. However, it does not work well for $d=1$ where $\theta_{max}(1)=0.748$ and $d=0$ where $\theta_{max}(0)=1$. To obtain better approximation, we added next order term in the nominator of relation (\ref{eq:fit1}):
\begin{equation}
\theta_{max}(d) = \frac{c_1 + c_2 d + c_3 d^2}{2^d}
\label{eq:fit2}
\end{equation}
Figure \ref{fig:qd16} presents plot of $\theta_{max}(d)$ where values for $1<d<2$ were obtained with (\ref{eq:qt}) and (\ref{eq:feder}) whereas values for $d \in \{2, 3, 4, 5, 6\}$ were taken from \cite{bib:Torquato2006}. 
\par
The fit defined by Eq.(\ref{eq:fit2}) matches the data for $d<2$ and is characterized by correlation coefficient $R=0.9992$. Furthermore, the parameter $c_1$ corresponds very well to random saturation density for $d=0$. The correction given by the third term $c_3 d^2$ = $0.086 \cdot d^2$ is important mostly for large $d$ values. Therefore, simulations for higher dimensions should determine whether $c_3$ should remain or not.
\subsection{Pair correlation function}
\begin{figure}[htb]
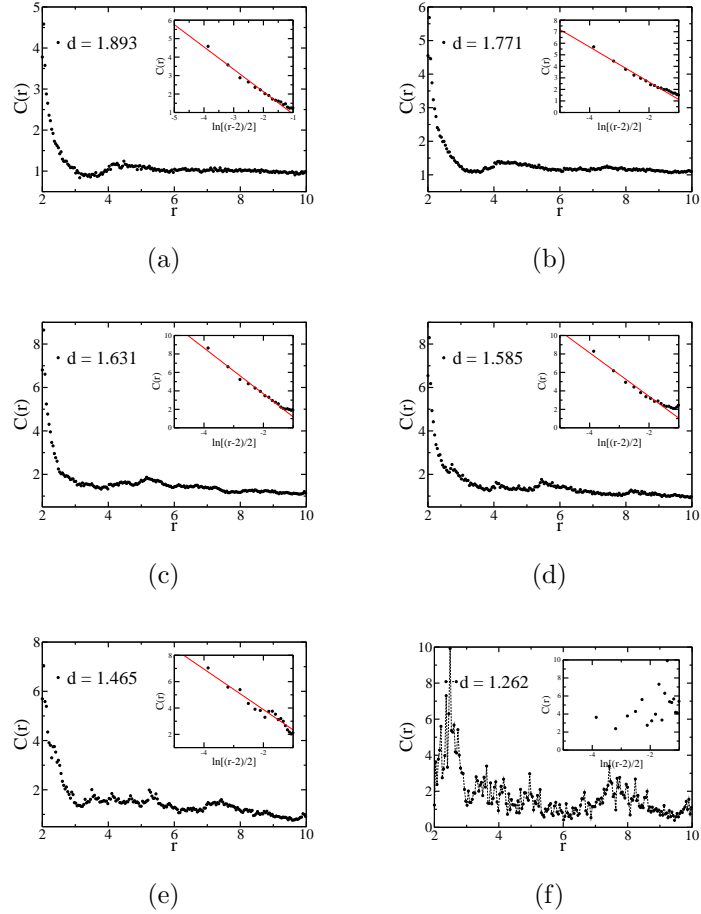

\vspace{0.1cm}
\centerline{%
\subfloat[]{\includegraphics[width=4cm]{cor495}}
\hspace{1cm}
\subfloat[]{\includegraphics[width=4cm]{cor381}}
}
\vspace{0.1cm}
\centerline{%
\subfloat[]{\includegraphics[width=4cm]{cor427}}
\hspace{1cm}
\subfloat[]{\includegraphics[width=4cm]{corst}}
}
\vspace{0.1cm}
\centerline{%
\subfloat[]{\includegraphics[width=4cm]{cor341}}
\hspace{1cm}
\subfloat[]{\includegraphics[width=4cm]{cor325}}
}
\caption{Two-point correlation function for collectors of different dimensions. Insets show asymptotic character for particles in close proximity to each other with a fit given by (\ref{eq:corlog}). (a)~$d=1.893$ (Sierpinski Carpet), (b)~$d=1.771$, (c)~$d=1.631$, (d)~$d=1.585$ (Sierpinski Triangle), (e)~$d=1.465$ (Vicsec fractal), (f)~$d=1.262$ (Cantor Dust).}
\label{fig:cor}
\end{figure}

Two-point correlation functions for fractal collectors are presented in Fig.\ref{fig:cor}.
Again, except for the case of Cantor Dust collector, all the plots look typical; for $d<1.7$, however, the specific fractal structure is manifested by a number of local extremes. Therefore, the pure superexponential decay (\ref{eq:corfast}) cannot be observed. For small values of $r$, the logarithmic singularity (\ref{eq:corlog}) appears just as for adsorption in higher dimensions. In case of the Cantor Dust collector, the two-point correlation function behaviour is strictly connected with cluster character of the fractal. There is also no point to investigate the singularities at kissing limit because close packing is restricted by incoherent structure of the fractal.
\subsection{Below d=1}
\begin{figure}[htb]
\vspace{1cm}
\centerline{%
\includegraphics[width=8cm]{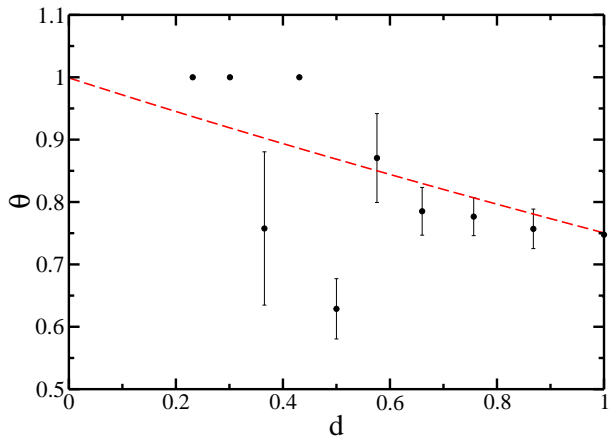}}
\caption{$\theta_{max}(d)$ for $d<1$. Dots represent data and dashed line is the (\ref{eq:fit2}) fit with $c_1=0.999$, $c_2=0.416$ and $c_3=0.086$ - the same as presented in Fig.\ref{fig:qd16}.}
\label{fig:qd06}
\end{figure}

The preliminary RSA simulation has been also performed for $d<1$ using General Cantor Set (GCS) as a collector. The GCS dimension depends on the size of middle part removed in each iteration:
\begin{equation}
d_{GCS}(q) = \frac{\ln(2)}{\ln[2/(1-q)]}
\end{equation}
where $q$ is the relative length of the removed fragment. For standard Cantor Set $q=1/3$ thus $d_{CS} = \ln(2) / \ln(3) \approx 0.631$. The diameter of the collector used during simulations was $100r_0$. The dependence of $\theta_{max}$ on dimension value is shown in Fig.\ref{fig:qd06}.
The data for $d>0.5$ are consistent with the fit (\ref{eq:fit2}) within an error range. For lower $d$, when holes in a collector are bigger, the geometrical relations start to play the most important role. The situation is somehow similar to the Cantor Dust case; however for this case, it can be proved that random maximal coverage ratio depends strictly on the particle diameter to collector size ratio. Therefore, the possibility of getting reliable data for $d<0.5$ from RSA simulation is doubtful. There is also no point in analysing the two-point correlation function because all collectors having $d<1$ are incoherent and the correlation behaviour will mainly express the structure of a specific collector. 
\section{Summary}
The RSA on fractal collectors having $d<2$ was studied. For such systems, the adsorption kinetics obey the Feder's law (\ref{eq:feder}). Moreover, the singularity of the two-point correlation function at kissing limit is logarithmic. The decay seems to be superexponential, but the autocorrelations for low dimensional fractals reflect also collector structure. It was one of the reasons why obtaining reliable results from simulation on collectors having dimension significantly lower than $d=1$ is not possible. The maximal random coverage ratio for the whole range of collector dimensions is governed by the Eq.(\ref{eq:fit2}), which is a simple extension to the rule (\ref{eq:fit1}) introduced and analytically supported for high dimensions.
\par
This work was supported by grant MNiSW/0013/H03/2010/70.


\section*{References}
\begin{thebibliography}{10}
\bibitem{bib:Kinge2008} S. Kinge, M. Crego-Calama, D.N. Reinhoudt, ChemPhysChem
{\bf 9} 1 20 (2008).
\bibitem{bib:Pfeifer1983} P. Pfeifer and D. Avnir, J. Chem. Phys. {\bf 79}, 3558 (1983)
\bibitem{bib:Avnir1983} D. Avnir, D. Farin, and P. Pfeifer, J. Chem. Phys. {\bf 79}, 3566 (1983); Surf. Sci. {\bf 126}, 569 (1983); Nature {\bf 30}, 261 (1984).
\bibitem{bib:Basillais1998} E. Basillais, Comptes Rendus de l'Academie des Sciences - Serie III, {\bf 321} (4), 295 (1998).
\bibitem{bib:Khasanov1991} M. M. Khasanov and I. I. Abyzbaev, J. Eng. Phys. Thermophys. {\bf 61} 6, 1516 (1991).
\bibitem{bib:Conway1998} J. H. Conway and N. J. A. Sloane, Sphere Packings, {\em Lattices and Groups}, Springer-Verlag, New York, 1998.
\bibitem{bib:Rogers1964} C. A. Rogers, {\em Packing and Covering}, Cambridge University Press, Cambridge, 1964.
\bibitem{bib:Privman2004} A.M.R. Cadilhe, V. Privman, Modern Phys. Lett. B {\bf 18}, 207 (2004).
\bibitem{bib:Fan1991} Y.Fan, J.K.Percus, Phys. Rev. Lett. {\bf 67} 13 1677 (1991).
\bibitem{bib:Nazzarro1996} M. S. Nazzarro, A. J. Ramirez Pastor, J. L. Riccardo and V. Pereyra, J. Phys. A: Math. Gen. {\bf 30} 1925 (1997).
\bibitem{bib:Loscar2003} E. S. Loscar, R. A. Borzi, and E. V. Albano, Phys. Rev. E {\bf 68} 041106 (2003).
\bibitem{bib:Cole1986} M. W. Cole and N. S. Holter, Phys. Rev. B {\bf 33} 8806 (1986).
\bibitem{bib:Pfeifer1989} P. Pfeifer, Y. J. Wu, M. W. Cole, J. Krim, Phys. Rev. Lett. {\bf 62} 1997 (1989).
\bibitem{bib:Brilliantov1998} N. V. Brilliantov, Yu. A. Andrienko, P. L. Krapivsky, and J. Kurths, Phys. Rev. E {\bf 58} 3530 (1998).
\bibitem{bib:Feder1980} J. Feder, J. Theor. Biol. 87, 237 (1980).
\bibitem{bib:Swendsen1981} R. H. Swendsen, Phys. Rev. A {\bf 24}, 504 (1981).
\bibitem{bib:Privman1991} V. Privman, J.-S. Wang, and P. Nielaba, Phys. Rev. B {\bf 43} 3366 (1991).
\bibitem{bib:Torquato2006} S. Torquato, O.U. Uche, F.H. Stillinger, Phys.Rev.E {\bf 74} 061308 (2006).
\bibitem{bib:Renyi1963} A. Reńyi, Sel. Trans. Math. Stat. Prob. {\bf 4}, 203 (1963).
Publ. Math. Inst. Hung. Acad. Sci. {\bf 3}, 109, (1958).
\bibitem{bib:Torquato2002} S. Torquato and F. H. Stillinger, J. Phys. Chem. B {\bf 106}, 8354 (2002).
\bibitem{bib:Bonnier1994} B. Bonnier, D. Boyer, and P. Viot, J. Phys. A {\bf 27}, 3671 (1994).
\bibitem{bib:Ciesla2012} M. Ciesla, J. Barbasz, J. Stat. Mech. 03015 (2012).
\bibitem{bib:Adamczyk2010} Z. Adamczyk, J. Barbasz, M. Ciesla, Langmuir, {\bf 26} 11934 (2010); Langmuir, {\bf 27} 6868 (2011).
\bibitem{bib:Ball1992} K. Ball, Int. Math. Res. Notices {\bf 68}, 217 (1992).

\end{thebibliography}
\end{document}